\documentclass{article}
\usepackage{amssymb}
\usepackage{amsfonts}
\usepackage{amsmath}
\usepackage{epsfig}
\usepackage{graphicx}

\begin{document}

\title{Long-range connections, real-world networks and rates of diffusion}
\author{\textit{Tanya Ara\'{u}jo}\thanks{%
tanya@iseg.ulisboa.pt} \\
%EndAName
UECE, ISEG, Universidade de Lisboa \and \textit{R. Vilela Mendes}\thanks{%
rvilela.mendes@gmail.com, rvmendes@fc.ul.pt, http://label2.ist.utl.pt/vilela/%
} \\
%EndAName
CMAFCIO, Universidade de Lisboa}
\date{ }
\maketitle

\begin{abstract}
Long range connections play an essential role in dynamical processes on networks, on the processing of information in biological networks, on the structure of social and economical networks and in the propagation of opinions and epidemics. Here we review the evidence for long range connections in real world networks and discuss the nature of the nonlocal diffusion arising from different distance-dependent laws. Particular attention is devoted to exponential and power laws.
\end{abstract}

\section{Introduction}

Long range connections play an important role in the dynamical processes on
networks. For example, the existence and relevance of long-range connections
in the brain has been studied \cite{Park} - \cite{Fluo} with diminished
long-range functional connectivity being associated to cognitive disorders 
\cite{Barttfeld}. Long-range connections are also important for the
structure of social and economic networks \cite{Hogan} \cite{Romantic} \cite%
{Carvalho} as well as for the propagation of epidemics \cite{Gustafson}.

Dynamics on networks involving jumps over many links or cascades of many
unit jumps, may lead to anomalous diffusion \cite{Riascos1} - \cite{Nigris}.
Likewise the existence of long range connections is expected to lead to
anomalous diffusion effects. When the density of long range connections
follows a power law, as a function of a suitably defined distance, the
propagation of signals in the network behaves as the solution of a
fractional diffusion equation \cite{Vilela2018} \cite{Vilela2020}.

Because networks with power-law connections, leading to superdiffusion,
display properties so very different from scale-free and hub dominated
networks, it has been proposed to classify them as a new network class, 
\textbf{the fractional networks\footnote{%
Networks with non-fractional couplings where nevertheless diffusion has
fractional features, not to be confused with fractionally coupled networks 
\cite{LiMa} \cite{Zuniga}.}}. However, it is to be expected that the nature
of the diffusion and therefore the propagation of information in the
network, may be different for other distance dependencies of the density of
long range connections. This is an important issue because, for example in
brain networks, existence of long-range connections between the specialized
modules does not guarantee global integration of the cognitive functions. It
is also necessary that the flow of information be sufficiently fast for the
stimulus integration to be performed in a timely manner\footnote{%
This seems of particular relevance for the forward and backwards loops in
the predictive coding mode of brain operation.}.

In Section 2 of this paper we review the empirical evidence for long range
connections in a few real-world networks. This includes both
distance-dependent results obtained by other authors in several networks as
well as our own analysis of some networks. A focus of our analysis is the
extraction of the distance-dependent functional laws of the connections. A
general result that is obtained in many networks is that short-distance and
long-distance connections follows different laws. This is consistent with
the modular nature of many networks, which have a high density of
connections inside local modules and a much smaller number of long-range
connections. This is typically the structure of brain networks. In these
networks the long-range connections between the modules favor integration of
perception but, on the other hand, they also imply a much larger biological
cost. As we will find out, this two-laws behavior also appears in a
transport network, again signaling a modular structure on the economic
organization of society. In social networks the law of short- and long-range
connections is more uniform. Probably it reflects the fact that short and
long range connections have a similar cost.

Section 3 contains a detailed study of the nature of nonlocal diffusion,
with particular emphasis on the comparison of the power law and exponential
distance-dependence of the connections. It is clear that all real-world
networks have a distance cut-off. Because of that, the distribution of the
connections intensity has compact support and it is well-known \cite{Rossi} 
\cite{Chasseigne} that, for compact support diffusion kernels, the
asymptotic behavior is identical to the one of the heat equation (normal
diffusion). However, what is important in practice is not the behavior of
the propagation of information for extremely large times, but for short and
intermediate large times. We discuss in detail this question in Section 3
for power laws with a cut-off and find out that indeed, for intermediate
large times, the solution behaves like an anomalous diffusion solution, very
different from the asymptotics for extremely large times. Analytical
estimates for small and large times are possible, but not for these
intermediate times of practical interest. By numerical calculations we have
nevertheless been able to parametrize and quantify the effective
"fractionality" of the diffusion for intermediate times.

Power-law and exponential distance-dependence of the connections lead to
very different diffusion behavior. However it is many times difficult to
distinguish between these two-dependencies from the experimental data. To
emphasize this point we have displayed for all analyzed networks, both
power-law and exponential fits. The reason for this difficulty is further
analyzed in Section 4 (see Fig.\ref{PROPAGA_2}). In Section 4 we also carry
out a numerical experiment of propagation of a unit signal in power-law,
exponential and nearest-neighbor networks. The smaller propagation times
are, of course, for the power-law networks, but even the exponential network
is much more efficient that the nearest-neighbor one.

The set of real-world networks that we were able to analyze is of course
limited. This was the data that we were able to obtain either from the Web
or kindly provided by the referenced authors. We encourage the readers that
have access to other data to pursue our preliminary work. Interesting
questions to explore are the nature of the functional law of connections in
the human brain (both structural and functional) as compared to other
mammals, as well as the nature of the "fractionality" index across species.
Are some brain impairments, that have been identified, related to the nature
of the distance-dependence law or just to the overall number of
long-distance connections? Data on the more popular social networks and
their distance-dependent links is harder to obtain. Interesting studies
might be performed in that data when available, correlated also with the
diffusion of trends and opinions. Some data on ecological and trophic
networks was available, but not sufficient for robust results. Notice that
in some of these networks the relevant notion of distance is not
geographical distance.

\section{Real-world networks: Analysis of empirical data}

\subsection{A human mobility network}

The first network that was analyzed was the network of flights to and from
airports in the New York region. 

\begin{figure}[htb]
\centering
\includegraphics[width=0.75\textwidth]{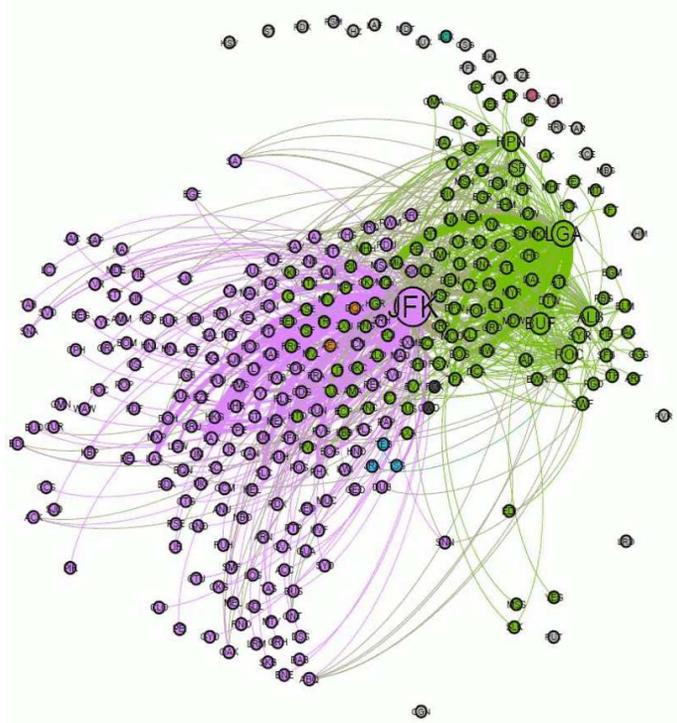}
\caption{The network of flights to or from New York airports in January 2019}
\label{airports}
\end{figure}

Data for the
month of January 2019, for example, was collected from the database
available in reference \cite{Transat2020}. The file contains the following
fields:

\# Passengers (number of passengers),

\# Distance (in miles),

\# Origin\_Airport\_ID (unique identifier Airport\_ID),

\# Destination\_Airport\_ID,

\# Month

From the data in these fields a weighted network of Airport\_IDs is defined.
The Airport\_IDs are the nodes while the strength of the links is given by
the number of passengers flying between the nodes (Origin\_Airport\_ID and
Destination\_Airport\_ID, or vice-versa) during the time interval being
considered (month), regardless of the flight direction. Figure \ref{airports}
shows the network of flights to and from New York airports in January 2019.
The network has 317 nodes (airports) and 641 weighted links, the strongest
link being a Delta Airline flight that links the airports of LaGuardia (LGA)
and Atlanta (ATL). The largest distance (10201 miles) concerns the JFK
airport in New York and the Melbourne Airport, in Australia. The average
degree is 4 and the average weighted degree is 98319. The network diameter
is 7 and the network modularity displays four different classes. The colors
in Figure \ref{airports} characterize each one of the modularity classes,
while the size of each node is proportional to its degree. The average
clustering coefficient is 0.49 and the average path length is 2.4.

\begin{figure}[htb]
\centering
\includegraphics[width=0.75\textwidth]{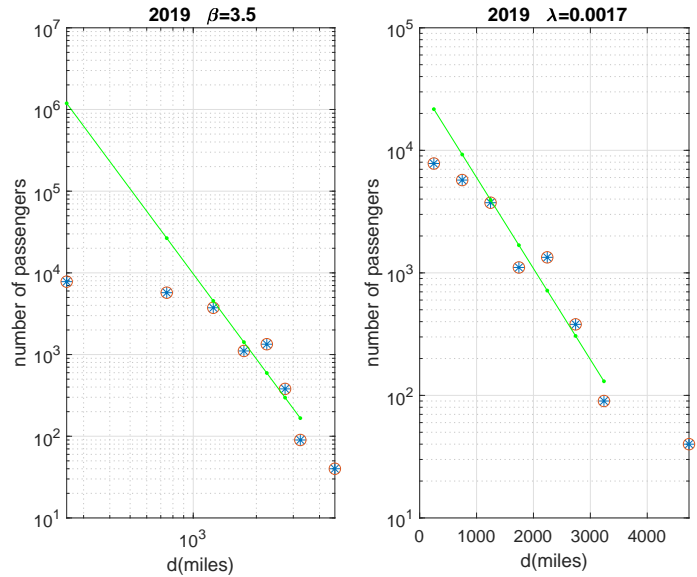}
\caption{Distance-dependent distribution
of the network links in the airports network for the year 2019. $\protect%
\beta $ is the exponent of the power law $w\sim d^{-\protect\beta }$ and $%
\protect\lambda $ the coefficient in the exponent of the exponential law $%
w\sim \exp \left( -\protect\lambda d\right) $.}
\label{aero19}
\end{figure}

\begin{figure}[htb]
\centering
\includegraphics[width=0.75\textwidth]{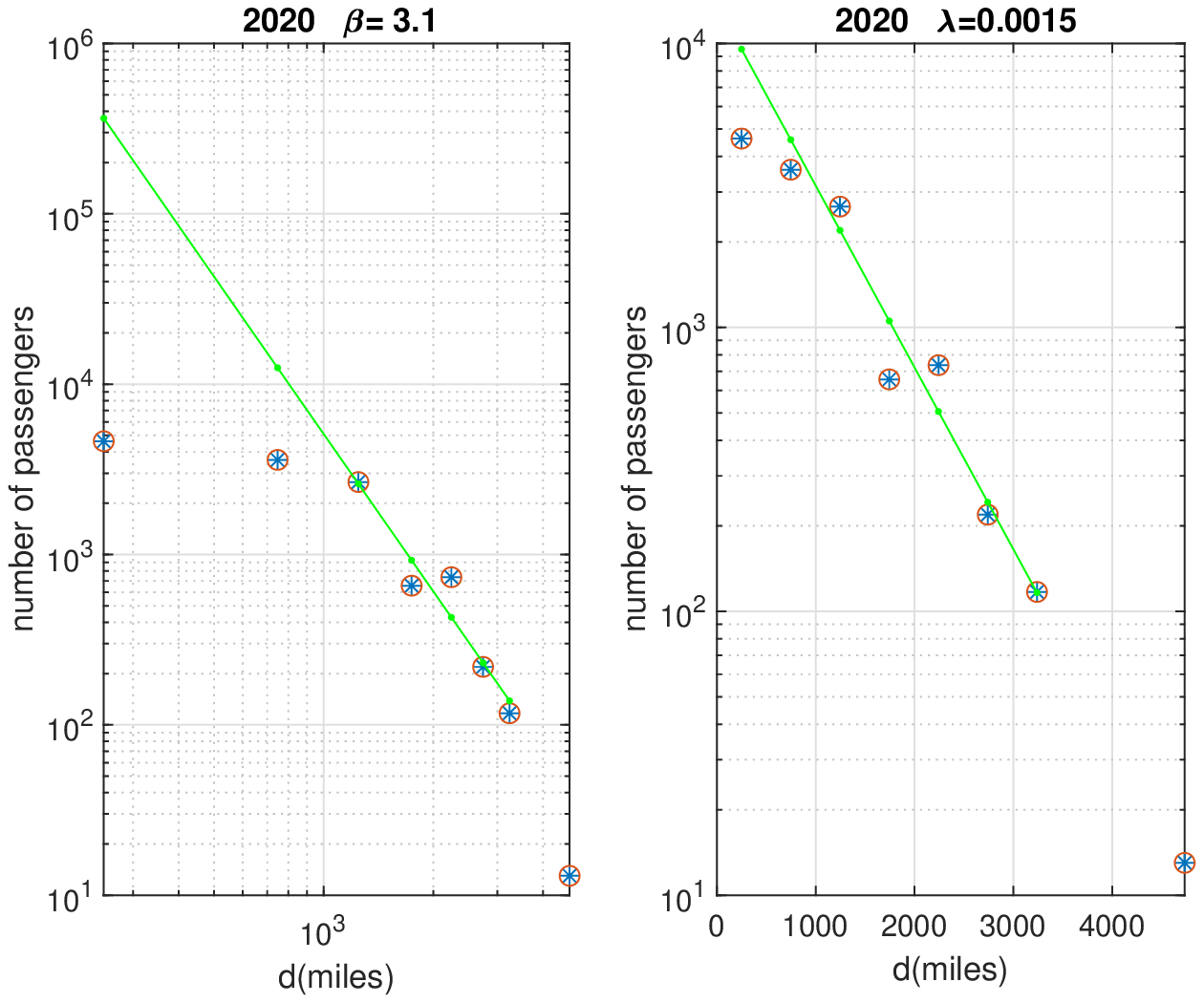}
\caption{Distance-dependent distribution of the network
links in the airports network for the year 2020. $\protect\beta $ is the
exponent of the power law $w\sim d^{-\protect\beta }$ and $\protect\lambda $
the coefficient in the exponent of the exponential law $w\sim \exp \left( -%
\protect\lambda d\right) $.}
\label{aero20}
\end{figure}

We have collected the same type of data for the full years of 2019 and 2020.
The results are plotted in Figs.\ref{aero19} and \ref{aero20}

In these figures we also make polynomial fits corresponding in the left-hand
plots to a power law $w\sim d^{-\beta }$ where $w$ is the number of
passengers and $d$ the inter-airports distance and to an exponential law $%
w\sim \exp \left( -\lambda d\right) $ in the right-hand plots. Our
conclusion is that the data is better described by a two-law structure
separating short and the long range connections, with the long range
connections obeying an approximate power law. According to the analysis in
Section 3, this power law leads to anomalous diffusion, which has relevance
for the fast propagation of trends and diseases.

Other authors have also emphasized the role of long-range connections in
human mobility networks. For example Viana and Costa \cite{Viana} analyzed
their role in the London urban and US highway networks mostly in connection
with the small world properties and travel velocity between nodes. The
statistics of travel patterns and returns to the same locations using cell
phones has also been studied \cite{Barabasi}. The conclusion being that the
individual travel patterns collapse into a single spatial probability
distribution, it would be interesting to find out whether there are
distinguishing features in short and long travel distances. . Another human
mobility that has been analyzed by Riascos and Mateos \cite{Mateos} involved
one billion taxi trips in New York City. The conclusion was that the
probability of a trip to a site inside a circular region of radius $R$
around the origin is approximately constant, whereas the probability to a
long range trip outside this circle decays as a power law with an
exponential cutoff. That is, a modular structure similar to what we have
found for the airports network.

\subsection{Brain networks}

Network theory is a valuable tool to explore the complex nature of brain
networks, the discovery of patterns of connections between cortical areas
being the focus of most research works (see for example \cite{Markov2013} 
\cite{Markov2014} \cite{Gamanut2017} \cite{Horvat2016}). Our analysis will
concentrate on macaque and mouse brain networks.

\subsubsection{Macaque}

The data used here was collected from Core-Net.org \cite{Markov2013}. In
this data the pathways linking cortical areas have been followed by
retrograde tracing experiments using injections of fluorescent retrograde
tracers\footnote{%
Localization of injection sites and labeled neurons was based on a new
reference atlas that includes 91 cortical areas mapped to the left
hemisphere of case M132\textit{\ }\cite{Markov2014}.}.
\begin{figure}[htb]
\centering
\includegraphics[width=0.75\textwidth]{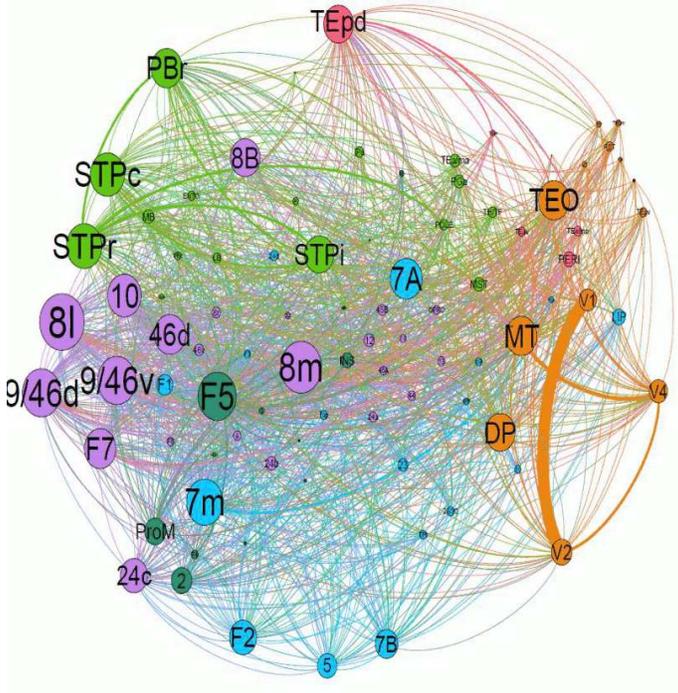}
\caption{The network of 91 brain areas collected from 39
macaques}
\label{macaque1}
\end{figure}

The file PNAS2013.xlsx downloaded from the data base has the following
fields: 1. Monkey, 2. Source area, 3. Target area, 4. Neurons, 5. FLNe
(fraction of labeled neurons), 6. Distance.

Source ($j$) and target ($i$) areas are labelled through retrograde labeling
using fluorescent tracers \cite{Horvat2016}. This labeling method reveals
all incoming connections $j_{i}$ to an injected (target) area $i$ by
labeling the cell bodies of the neurons in source area $j$ for which their
axons connect to area $i$. The fraction of labeled neurons (FLNe), given by
the ratio $FLNe=\frac{Neurons}{TotalNeurons}$ , may be interpreted as the
probability of a neuronal projection from Source area $i$ to Target area $j$ 
\cite{Ercsey2013}.

The fields Source and Target provide a weighted network of 91 brain areas.
The Source and Target areas are the nodes while the intensity of the links
corresponds to the number of Neurons connecting these areas. Figure \ref%
{macaque1} shows the network of the 91 areas in the PNAS2013.xlsx file, 39
macaques being sampled in 1989 instances of the database. Thus, this macaque
network has 91 nodes, connected by 1401 links, the size of each node being
proportional to its degree. The average degree is 30.7 and the average
weighted degree is 869967.3. The network diameter is 3 and the network
modularity displays six different classes, characterized by different colors
in the figure. The average clustering coefficient is 0.74 and the average
path length is 1.6.

\begin{figure}[htb]
\centering
\includegraphics[width=0.75\textwidth]{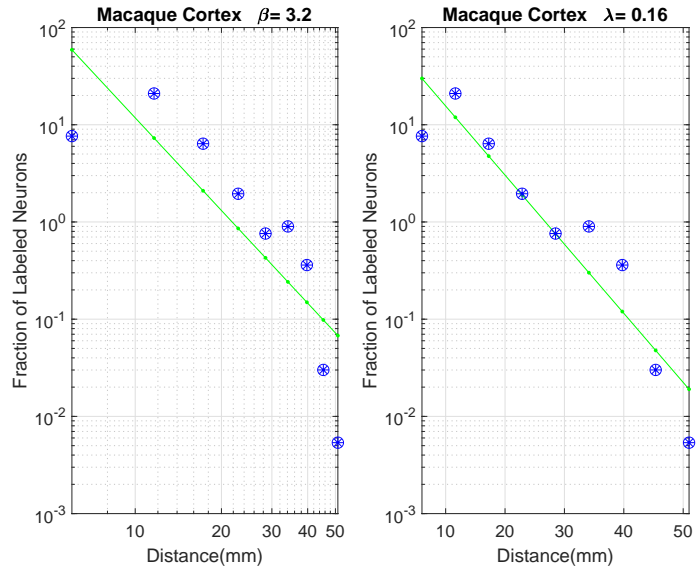}
\caption{Distance-dependent distribution of the network links for
the macaque cortex. $\protect\beta $ and $\protect\lambda $ have the same
meaning as before.}
\label{macaque2}
\end{figure}

The distance-dependent distribution of the network links is shown in Figure %
\ref{macaque2} both in a log-log and a semilog plot.

It seems that in this
case the better fit is an exponential one $w=exp({-\lambda }d)$, $w$ being
the links intensity (number of neurons) and $\lambda =0.16mm^{-1}$.

Likewise, if instead of the Neuron field we focus on the Fraction of Labeled
Neurons (FLNe) the result is quite similar, being in accordance with the
work in \cite{Horvat2016}, which by using data from macaque and mouse,
infers the existence of a general organizational principle based on an
exponential distance rule

\begin{equation}
FLNe(d)=c\ast exp({-\lambda d})  \label{1}
\end{equation}

\subsubsection{Mouse}

The database used here is obtained from reference \cite{Gamanut2017},
MouseDatabase.xlsx. The file MouseDatabase.xlsx contains, among others, the
following fields:

\begin{figure}[htb]
\centering
\includegraphics[width=0.75\textwidth]{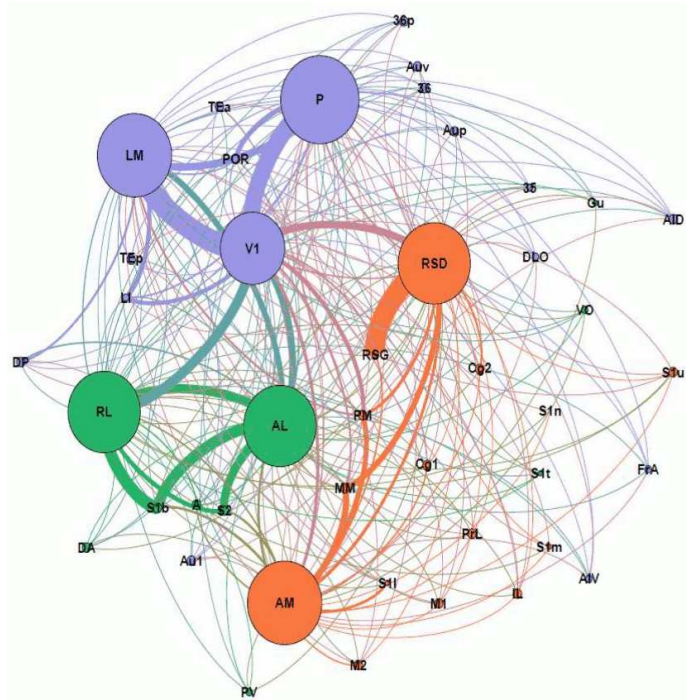}
\caption{The network of
neurons in 47 brain areas of 27 mice}
\label{mouse1}
\end{figure}

1. Mouse, 2. Source
area, 3. Target area, 4. FLNe (fraction of labeled neurons), 5. Neurons, 6.
Distance. The fields Source and Target allow to define a network of 47 brain
areas. The Source and Target areas are the nodes while the links correspond
to the number of Neurons between the Source and Target areas, as in the
macaque example studied before. Figure \ref{mouse1} shows the network of 47
areas in MouseDatabase.xlsx, where 27 mice were sampled in 1242 instances of
the database. The mouse network has 47 nodes and 703 weighted links.

In this mouse example, the number of target areas is much smaller than in
the macaque one. There are just seven (AM, AL, RL, P, ML, RDS and V1) target
areas (against 43 source ones) and therefore these are the areas with the
largest degrees, as Figure \ref{mouse1} shows. The network diameter is 2 and
the network modularity displays three different classes characterized by
different colors in Figure \ref{mouse1}. The average clustering coefficient
is 0.8 and the average path length is 1.35.
\begin{figure}[htb]
\centering
\includegraphics[width=0.75\textwidth]{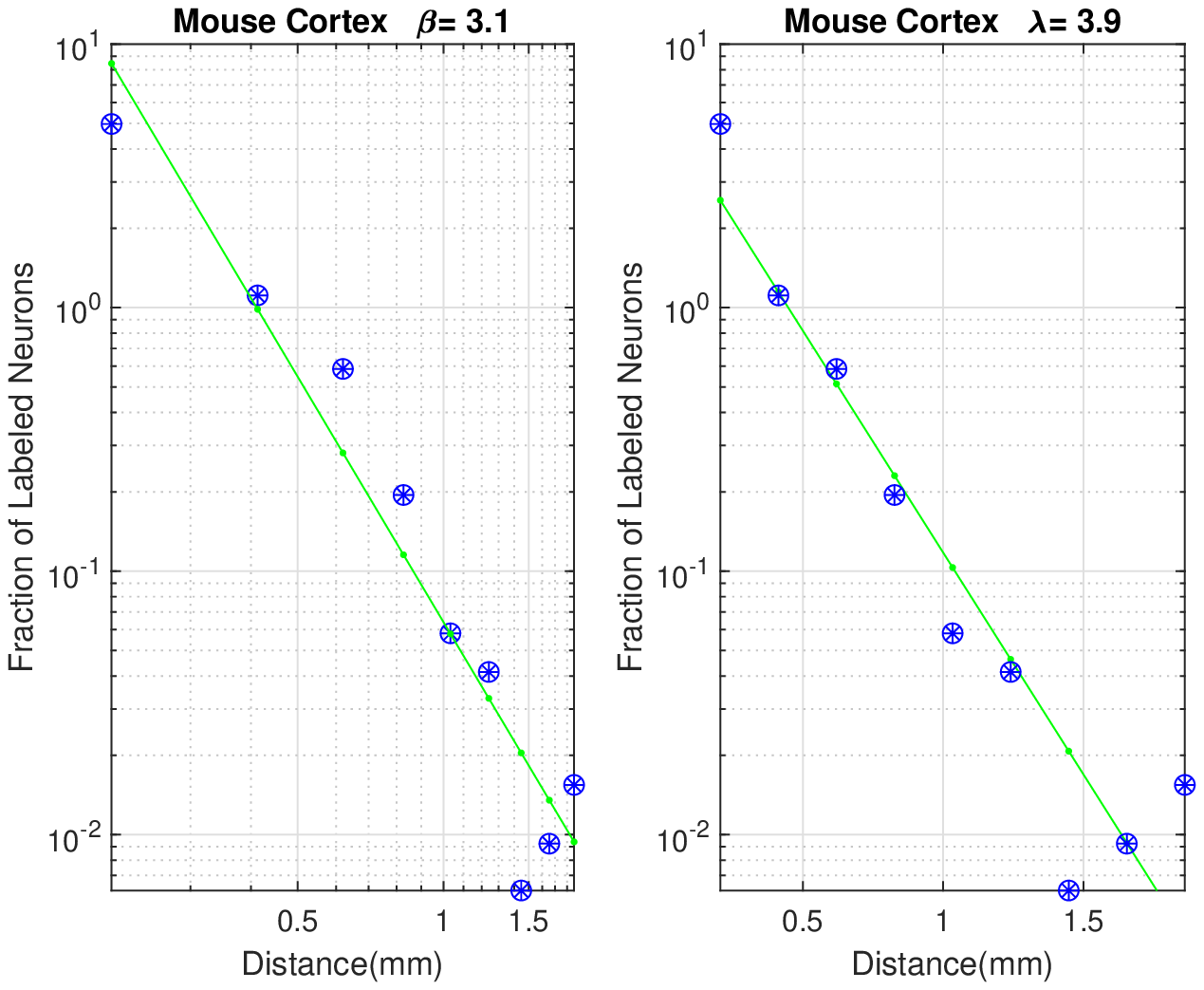}
\caption{Distance-dependent distribution of the links in the mouse
network. $\protect\beta $ and $\protect\lambda $ have the same meaning as
before.}
\label{mouse2}
\end{figure}

The distance-dependent distribution of the links is shown in Figure \ref%
{mouse2}. This is a case where it is difficult to decide between an
exponential and a power law. For an exponential law we would have $w\sim exp(%
{-\lambda }d)$ with $\lambda =3.9mm^{-1}$. The average distance in our
sample is $<d>_{mouse}=0.78mm$, smaller than the one reported in the samples
analyzed by other authors (for example $<d>_{mouse}=4.54mm$ in \cite%
{Gamanut2017}). The occurrence of an exponential law for the macaque and
mouse brain networks is in fact favored by most authors. Notice however that
Knox et al. \cite{Knox} in a high resolution study of the mouse connectome
shows a power-law fit to the normalized connection density as a function of
the distance. The statistics of the data fit is however too poor to be able
to decide. It would be interesting to compare the functional laws in the
macaque and mouse networks with those of the human brain, both structural
and functional, for which we have not yet had access to reliable data.

\subsection{A social network}

Here we analyze the Brightkite location-based online social network \cite%
{Ryan}. Brightkite is a location-based social networking service provider
where users share their locations by checking-in. This friendship network
was collected using users public API, and consists of $58,228$ nodes and $%
214,078$ edges. The network is originally directed but later transformed
into a network with undirected edges when there is a friendship in both
ways. Data include a total of $4,491,143$ checkins of the network users over
the period April 2008 - October 2010.

The first downloaded file (Loc-brightkite\_totalCheckins.txt) with
information on the users location contains the fields: 1. User, 2. Latitude,
3. Longitude, 4. Location\_id, 5. Check-in time. The second Brightkite file
(loc-brightkite\_edges.txt) is a friendship network of Brightkite users ($%
58,228$ nodes and $214,078$ edges).

From the locations of each pair of Brightkite friends the corresponding
distance is computed. The characterization of the distance-dependent
distribution of the network links (number of friends) is shown in Figure \ref%
{Brightkite}.
\begin{figure}[htb]
\centering
\includegraphics[width=0.75\textwidth]{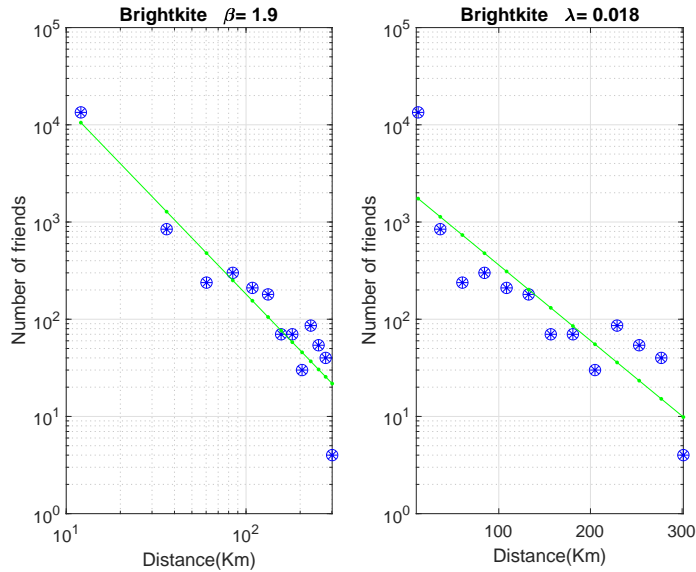}
\caption{Friends vs.
distance in the Brightkite network. $\protect\beta $ and $\protect\lambda $
have the same meaning as before.}
\label{Brightkite}
\end{figure}
Figure \ref%
{Brightkite} shows a polynomial fits (in log-log and semilog plots) of the
number of friends $w_{i,j}$ and their distances $d_{ij}$. The best fit
corresponds to a power laws $w_{ij}=d_{ij}^{-\beta }$ with $\beta =1.9$. In
this case one obtains a fairly uniform power law without any indication of a
modular structure.

\subsection{A fungi network}

The network studied here is part of a large data set of $269$ fungal
networks available at \cite{Fungi1} \cite{Fungi2}. The data providers
construct fungal networks by estimating cord conductances. In defining
fungal networks, the nodes are located at hyphal tips, branch points, and
anastomoses, while the edges represent cords. Structural networks are
constructed by calculating edge weights based on how much nutrient traffic
is predicted to occur along each edge.

\begin{figure}[htb]
\centering
\includegraphics[width=0.75\textwidth]{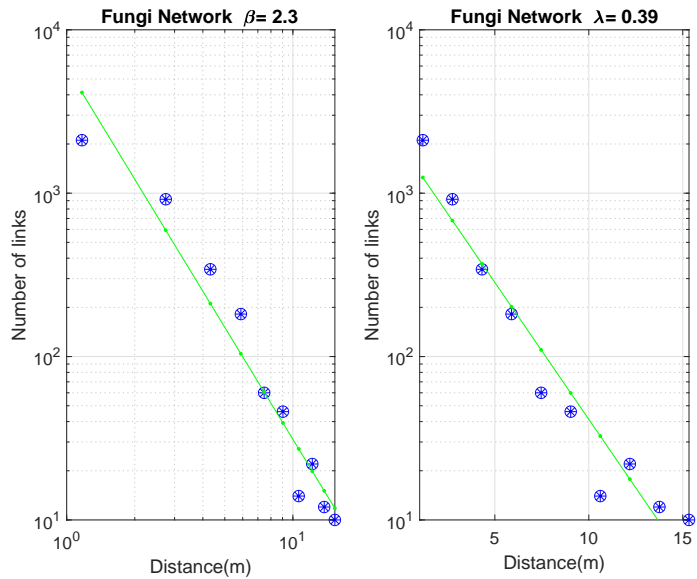}
\caption{A fungi network.$\protect\beta $ and $\protect\lambda $
have the same meaning as before.}
\label{Fungi}
\end{figure}

Being location-based networks, in addition to the complete list of network
links, the data set also includes the coordinates of each node. For the
network present here, we used the sample identified as $Pp\_M\_Tokyo\_U\_N%
\_26h\_1$. This sample has $1357$ nodes and $3716$ undirect links. From the
coordinates of each pair of linked nodes the corresponding distance is
computed. The characterization of the distance-dependent distribution of the
network links is shown in Figure \ref{Fungi}. 

This is again a
situation where it is difficult to decide whether to infer an exponential or
a power-law.

\section{Nonlocal diffusion and long range network connections}

Short and long-range connections determine the global behavior of a network,
both its emergent\ dynamical evolution and the spread of information (or
pathologies) throughout the network. It is intuitive that long-range
connections must play a role on the speed of the spread of information as
well as on the context integration of the sensory inputs in brain networks.
However it is not so obvious to infer how such phenomena may depend on the
density and distance-dependence of the long-range connections.

Here we make an attempt to address these questions in the framework of the
theory of nonlocal diffusion \cite{Rossi} \cite{Bucur} \cite{Vazquez}. Let
some time-dependent quantity (a field) $\phi \left( t,x\right) $ stand for
the density of individuals, quantity of information or activation in a $d-$%
dimensional space $X$. There are two related problems of nonlocal diffusion.
If $\phi \left( t,x\right) $ stands for the density of individuals or the
degree of activation, the nonlocal diffusion equation is 
\begin{equation}
\frac{\partial }{\partial t}\phi \left( t,x\right) =\int_{X}\rho \left(
y,x\right) \phi \left( t,y\right) d^{n}y-\phi \left( t,x\right) ,  \label{N1}
\end{equation}%
$\rho \left( y,x\right) $ being the probability density for a jump (or
transmission) from $y$ to $x$ and the last term accounts for the rate of
departure from $x$ to other locations.%
\begin{equation}
-\phi \left( t,x\right) =-\int_{X}\rho \left( x,y\right) \phi \left(
t,x\right) d^{n}y  \label{N2}
\end{equation}%
because $\int_{X}\rho \left( x,y\right) d^{n}y=1$. Eq.(\ref{N1}) conserves
the quantity $Q=\int_{X}\phi \left( t,x\right) d^{n}x$. In addition one
might consider additional terms added to the right hand side of these
equations to represent nonlocal internal interactions, localized sources or
consumption terms.

However if $\phi \left( t,x\right) $ stands for an information quantity or a
disease, the last term might not make sense. Information does not decay in
transmission to other nodes nor a disease is cured by infection of the
neighbors. To simply suppress the last term in Eq.(\ref{N1}) to describe
this situation does not make sense either because then $Q$ grows
exponentially. We will come back to this question on our numerical
simulations of propagation of an impulse of information and for the time
being our analysis will concern the Eq.(\ref{N1}).

In the continuous approximation, which would be a good approximation for a
network with a very large number of nodes, the kernel $\rho \left(
y,x\right) $ may be considered to be proportional to the density of
connections between nodes at the positions $y$ and $x$. Here we will be
concerned with the case where the kernel $\rho \left( y,x\right) $ is a
function of the distance $\left\vert y-x\right\vert $ only%
\[
\rho \left( y,x\right) =\rho \left( \left\vert y-x\right\vert \right) , 
\]%
with the distance defined by a problem-adapted metric. Without loss of
generality we will assume the parametrization of the network to be such that
the distances are Euclidean distances on that parametrization.

From our analysis of the real-world networks with long-range connections we
have concluded that two important functional distance dependencies are the
power law and the exponential law. It is therefore for these two laws that
our calculations will be performed. As for dimensionality, two most
important cases are $d=2$ and $d=3$. However other higher dimensions may be
important as well, when the network nodes are characterized by many
parameters. Important examples are social networks, credit scoring networks,
etc. Therefore we will derive general results for arbitrary (finite)
dimensions. Our main interest is the asymptotic behavior of $\phi \left(
t,x\right) $ for long and intermediate times. For $\rho \left( y,x\right)
=\rho \left( y-x\right) $ the integral in (\ref{N1}) is a convolution and a
Fourier transform treatment is appropriate%
\begin{equation}
\frac{\partial }{\partial t}\widetilde{\phi }\left( t,k\right) =\widetilde{%
\rho }\left( k\right) \widetilde{\phi }\left( t,k\right) -\widetilde{\phi }%
\left( t,k\right)  \label{N3}
\end{equation}%
with%
\begin{eqnarray}
\widetilde{\phi }\left( t,k\right) &=&\int_{X}e^{ik\cdot x}\phi \left(
t,x\right) d^{n}x  \nonumber \\
\widetilde{\rho }\left( k\right) &=&\int_{X}e^{ik\cdot x}\rho \left(
y-x\right) d^{n}x.  \label{N4}
\end{eqnarray}%
From Eq.(\ref{N3})%
\begin{eqnarray}
\widetilde{\phi }\left( t,k\right) &=&\widetilde{\phi }\left( 0,k\right)
\exp \left( t\left( \widetilde{\rho }\left( k\right) -1\right) \right) 
\nonumber \\
\phi \left( t,x\right) &=&\frac{1}{\left( 2\pi \right) ^{n}}\int_{\widetilde{%
X}}e^{-ik\cdot x}\widetilde{\phi }\left( t,k\right) d^{n}k  \label{N5}
\end{eqnarray}%
For definiteness we consider a Cauchy problem corresponding to the diffusion
of a unit pulse at the origin at time zero. That is, we are considering what
is called the fundamental solution to the equation. Once this is found,
solutions for arbitrary smooth initial conditions are obtained by
convolution with the fundamental solution. With, 
\begin{equation}
\widetilde{\phi }\left( 0,k\right) =1,  \label{N6}
\end{equation}%
the solution is%
\begin{equation}
\phi \left( t,x\right) =\frac{1}{\left( 2\pi \right) ^{n}}\int_{\widetilde{X}%
}d^{n}ke^{-ik\cdot x}\exp \left( \widetilde{\rho }\left( k\right) t\right)
\label{N6a}
\end{equation}%
with, for a kernel which only depends on the modulus of the distance ($\rho
\left( \left\vert y-x\right\vert \right) $), in dimension $n$%
\begin{equation}
\widetilde{\rho }\left( k\right) =\widetilde{\rho }\left( \left\vert
k\right\vert \right) =\left( 2\pi \right) ^{n/2}\int drr^{n-1}\rho \left(
r\right) \frac{J_{\frac{n}{2}-1}\left( \left\vert k\right\vert r\right) }{%
\left( \left\vert k\right\vert r\right) ^{\frac{n}{2}-1}},  \label{N7}
\end{equation}

We will now particularize this solution for two types of kernels:

\subsection{The power law case}

Let the connection strength be proportional to $\frac{c_{1}}{r^{\beta }}$
for most of the range of $r$. However, because this function is not
normalizable, one truncates it with a $G\left( r\right) $ function%
\[
G\left( r\right) =\left\{ 
\begin{array}{l}
1\text{ for }r_{\min }\leq r\leq r_{\max } \\ 
0\text{ otherwise}%
\end{array}%
\right. 
\]%
\begin{equation}
\rho _{1}\left( r\right) =\frac{c_{1}}{r^{\beta }}G\left( r\right) ,
\label{N9}
\end{equation}%
the normalization, $\int d^{n}x\rho _{1}\left( r\right) =1$, implying%
\begin{equation}
c_{1}=\left\{ 
\begin{array}{ccc}
\frac{\Gamma \left( n/2\right) }{2\pi ^{n/2}}\frac{n-\beta }{\left( r_{\max
}^{n-\beta }-r_{\min }^{n-\beta }\right) } &  & \left( n\neq \beta \right)
\\ 
\frac{\Gamma \left( n/2\right) }{2\pi ^{n/2}}\left( \log \frac{r_{\max }}{%
r_{\min }}\right) ^{-1} &  & \left( n=\beta \right)%
\end{array}%
\right.  \label{N10}
\end{equation}%
The asymptotic behavior for large times of the solution of Eq.(\ref{N1}) and
its relation to the behavior of the Fourier transform $\widetilde{\rho }%
_{1}\left( \left\vert k\right\vert \right) $ of $\rho _{1}\left( r\right) $
has been discussed by several authors in the past (see for example \cite%
{Rossi} \cite{Chasseigne}). Namely, the behavior of the solution for large
times is controlled by the functional dependence of $\widetilde{\rho }%
_{1}\left( \left\vert k\right\vert \right) $ at small $\left\vert
k\right\vert $. In particular, if for small $\left\vert k\right\vert $%
\begin{equation}
\widetilde{\rho }_{1}\left( \left\vert k\right\vert \right) =1-A\left\vert
k\right\vert ^{\alpha }+o\left( \left\vert k\right\vert ^{\alpha }\right)
\label{N11}
\end{equation}%
then the large time asymptotic behavior of the solution is the same as for
the solution $v\left( t,x\right) $ of the fractional Laplacian equation $%
\partial _{t}v(t,x)=-A(-\Delta )^{\alpha /2}v(t,x)$ with the same initial
condition, namely%
\begin{equation}
\lim_{t\rightarrow \infty }t^{n/\alpha }\max_{x}\left\{ \phi \left(
t,x\right) -v(t,x)\right\} =0  \label{N12}
\end{equation}%
with an $L^{\infty }$ rate of decay of the solution $t^{-n/\alpha }$. In
particular for the $\alpha =2$ case, the asymptotic behavior is identical to
the heat equation with a decay rate $t^{-n/2}$ and the asymptotic profile is
a Gaussian. Furthermore, refined asymptotics in terms of the derivatives of
the fundamental solution, identical to those of the heat equation, were
obtained \cite{Ignat-1}. What in principle would be of special importance
for our network setting is also the fact that the same kind of estimate is
obtained for nonlocal diffusion on a lattice \cite{Ignat-2}. All these are
very interesting mathematical results, which however may not be very useful
at relatively large, but finite, times. Let us compute the small $\left\vert
k\right\vert $ behavior of the power law Fourier transform $\widetilde{\rho }%
_{1}\left( \left\vert k\right\vert \right) $,%
\begin{eqnarray}
\widetilde{\rho }_{1}\left( \left\vert k\right\vert \right) &=&\left( 2\pi
\right) ^{\frac{n}{2}}c_{1}\int_{r_{\min }}^{r_{\max }}drr^{n-1}\frac{J_{%
\frac{n}{2}-1}\left( \left\vert k\right\vert r\right) }{\left( \left\vert
k\right\vert r\right) ^{\frac{n}{2}-1}}\frac{1}{r^{\beta }}  \nonumber \\
&=&1-A\left\vert k\right\vert ^{2}+\cdots  \label{N13}
\end{eqnarray}%
with%
\begin{equation}
A=\left\{ 
\begin{array}{llll}
\frac{n-\beta }{2n\left( n+2-\beta \right) }\frac{r_{\max }^{n+2-\beta
}-r_{\min }^{n+2-\beta }}{r_{\max }^{n-\beta }-r_{\min }^{n-\beta }} &  & 
\text{for} & n\neq \beta \\ 
\frac{1}{4n}\frac{r_{\max }^{2}-r_{\min }^{2}}{\log r_{\max }-\log r_{\min }}
&  & \text{for} & n=\beta%
\end{array}%
\right.  \label{N14}
\end{equation}%
According to the large time asymptotic results quoted above, the $\alpha =2$
behavior of the small $\left\vert k\right\vert $ behavior of $\widetilde{%
\rho }_{1}\left( \left\vert k\right\vert \right) $, might lead us to expect
for this case a diffusion similar to the heat equation. This may indeed be
true for extremely large times but not necessarily for the small or even
relatively large times of practical importance. In any case the $\alpha =2$
was to be expected, of course, because the $G\left( r\right) $ truncation
makes $\rho _{1}\left( r\right) $ a compact support function \cite%
{Chasseigne}.

\begin{figure}[htb]
\centering
\includegraphics[width=0.75\textwidth]{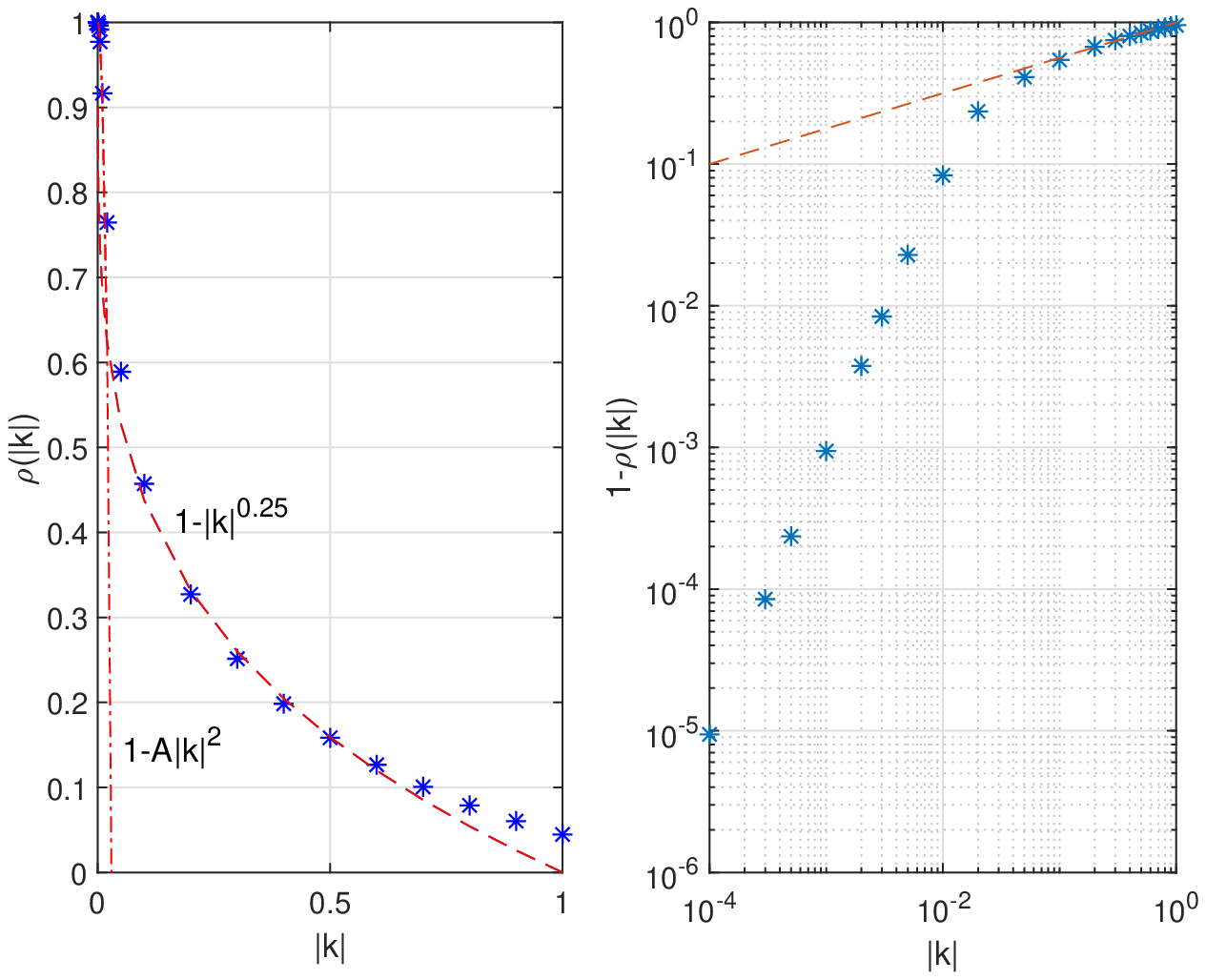}
\caption{$\protect\widetilde{\protect\rho }\left(
\left\vert k\right\vert \right) $ for the power law case with $\protect\beta %
=2$, $n=2$, $r_{\min }=1$, $r_{\max }=200$}
\label{SCALE}
\end{figure}

The limited usefulness of the (\ref{N13}) expansion is already apparent from
the fact that whenever the power law range is large (large $\left\vert
r_{\max }-r_{\min }\right\vert $) the $A$ coefficient becomes very large.
Then $\widetilde{\rho }_{1}\left( \left\vert k\right\vert \right) $ decays
with this rate only for a tiny fraction of $\left\vert k\right\vert $ which
contributes little to the inverse Fourier transform $\phi \left( t,x\right) $
in (\ref{N6a}). We illustrate this fact by numerically computing $\widetilde{%
\rho }_{1}\left( \left\vert k\right\vert \right) $ in the $n=2$, $\beta =2$
\ case with $r_{\min }=1$ and $r_{\max }=200$ (Fig.\ref{SCALE}).

One sees that the $\alpha =2$ scaling only occurs for very small $%
\left\vert k\right\vert $. This is put in evidence by plotting in the same
figure the line $1-A\left\vert k\right\vert ^{2}$. At around $\left\vert
k\right\vert =0.02$ there is a break for a quite different slope. We have
numerically computed the slop $\alpha $ after the break for dimensions $n=2$
to $n=4$ and several $\beta $ values. The results are plotted in Fig.\ref%
{SLOPES} (left panel) which shows that they seem to follow a universal
function of $\left( n-\beta \right) $. A rough fitting to a sigmoid function
leads to $\alpha (n-\beta )\sim 2/(1+\exp \left( \gamma \left( n-\beta
+c\right) \right) )$ with $\gamma =1.5$ and $c=1.1$.

\begin{figure}[htb]
\centering
\includegraphics[width=0.75\textwidth]{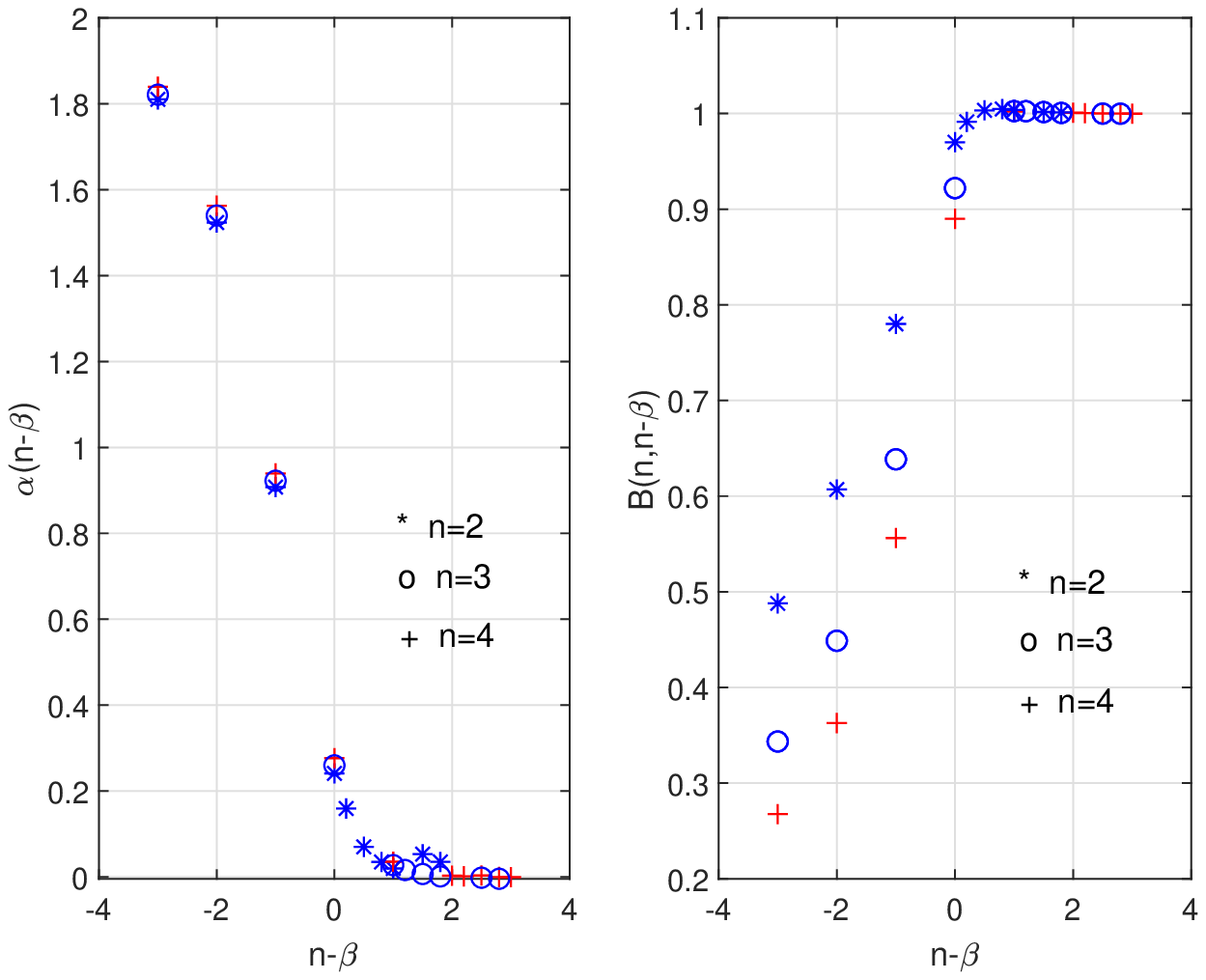}
\caption{The power exponent of the function $1-\protect\widetilde{%
\protect\rho }_{1}\left( \left\vert k\right\vert \right) $ after the break
and the $B$ coefficient in the $\protect\widetilde{\protect\rho }_{1}\left(
\left\vert k\right\vert \right) \simeq 1-B\left\vert k\right\vert ^{\protect%
\alpha }$ approximation}
\label{SLOPES}
\end{figure}
In the right-hand side panel we have plotted the factor $B$ in the 
$\widetilde{\rho }_{1}\left( \left\vert k\right\vert \right) \simeq
1-B\left\vert k\right\vert ^{\alpha }$ approximation with $\alpha $ as in
the left hand panel. The $B$ coefficient has a weak dependence on $n$.

One sees that for large negative $n-\beta $ the power exponent is consistent
with normal diffusion, but there is a large range of $n-\beta $
corresponding to anomalous superdiffusion. This applies to intermediate
times whereas, of course, for extremely large times the compact support
nature of $\rho \left( r\right) $ restores the slope to the one of normal
diffusion $\left( \alpha =2\right) $. As found above the dependence of the
intermediate scaling factor is far more complex than the one guessed from
comparison with the Grunwald-Letnikov representation of the fractional
derivative \cite{Vilela2020}. In conclusion:

For power-law powers $\beta $, such that $n-\beta \geq -1$ the nonlocal
equation will display for intermediate times a fractional superdiffusive
behavior with exponent that follows an universal law as represented in Fig.%
\ref{SLOPES}. In Fig.\ref{SCALE} we have also plotted the function $%
1-\left\vert k\right\vert ^{\frac{1}{4}}$ to emphasize how this is a better
first order approximation as compared to the asymptotic $1-A\left\vert
k\right\vert ^{2}$.

\begin{figure}[htb]
\centering
\includegraphics[width=0.75\textwidth]{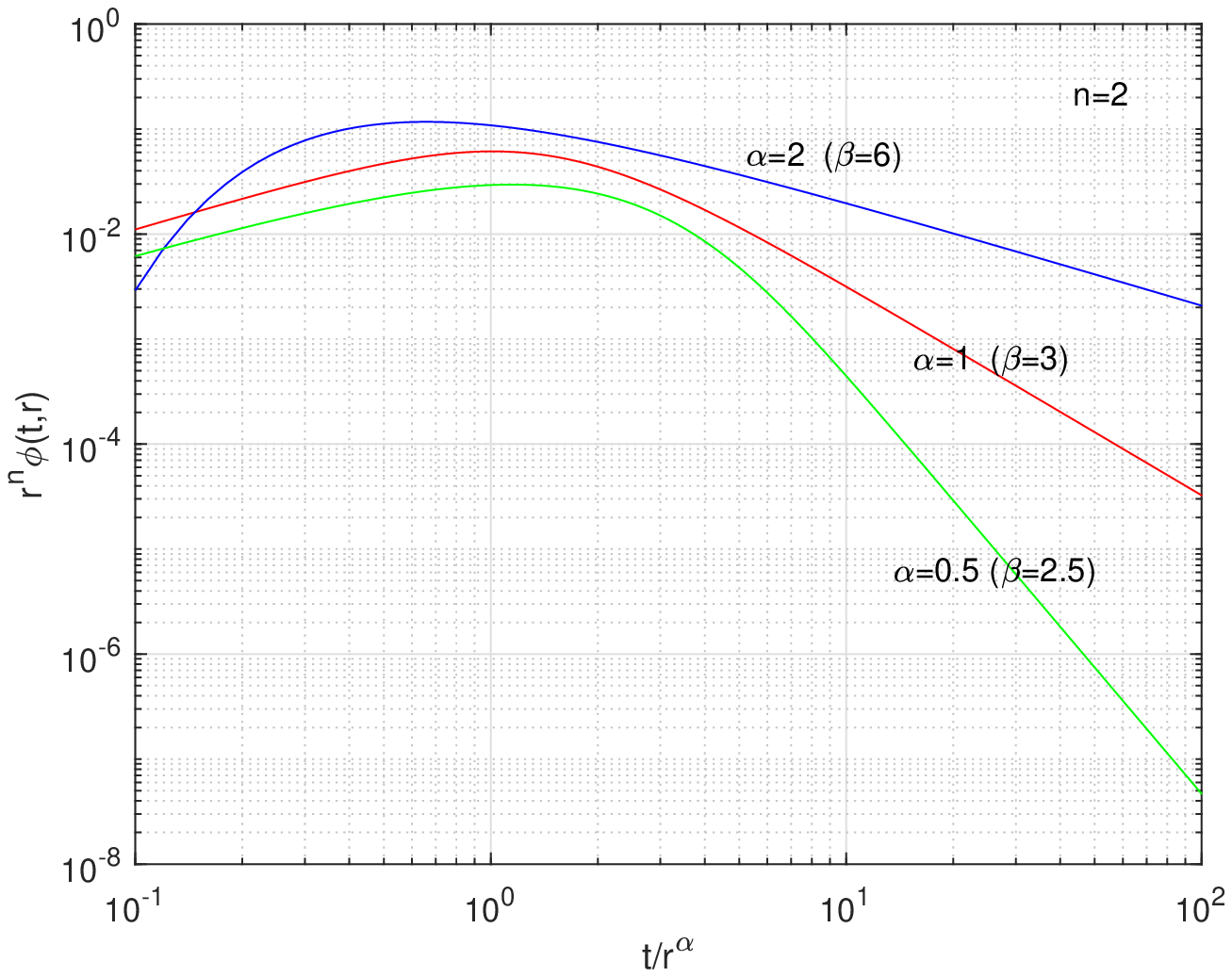}
\caption{The $r^{n}\protect\phi \left(
t,r\right) $ function in dimension $2$ for the power law case}
\label{PHI_2}
\end{figure}

\begin{figure}[htb]
\centering
\includegraphics[width=0.75\textwidth]{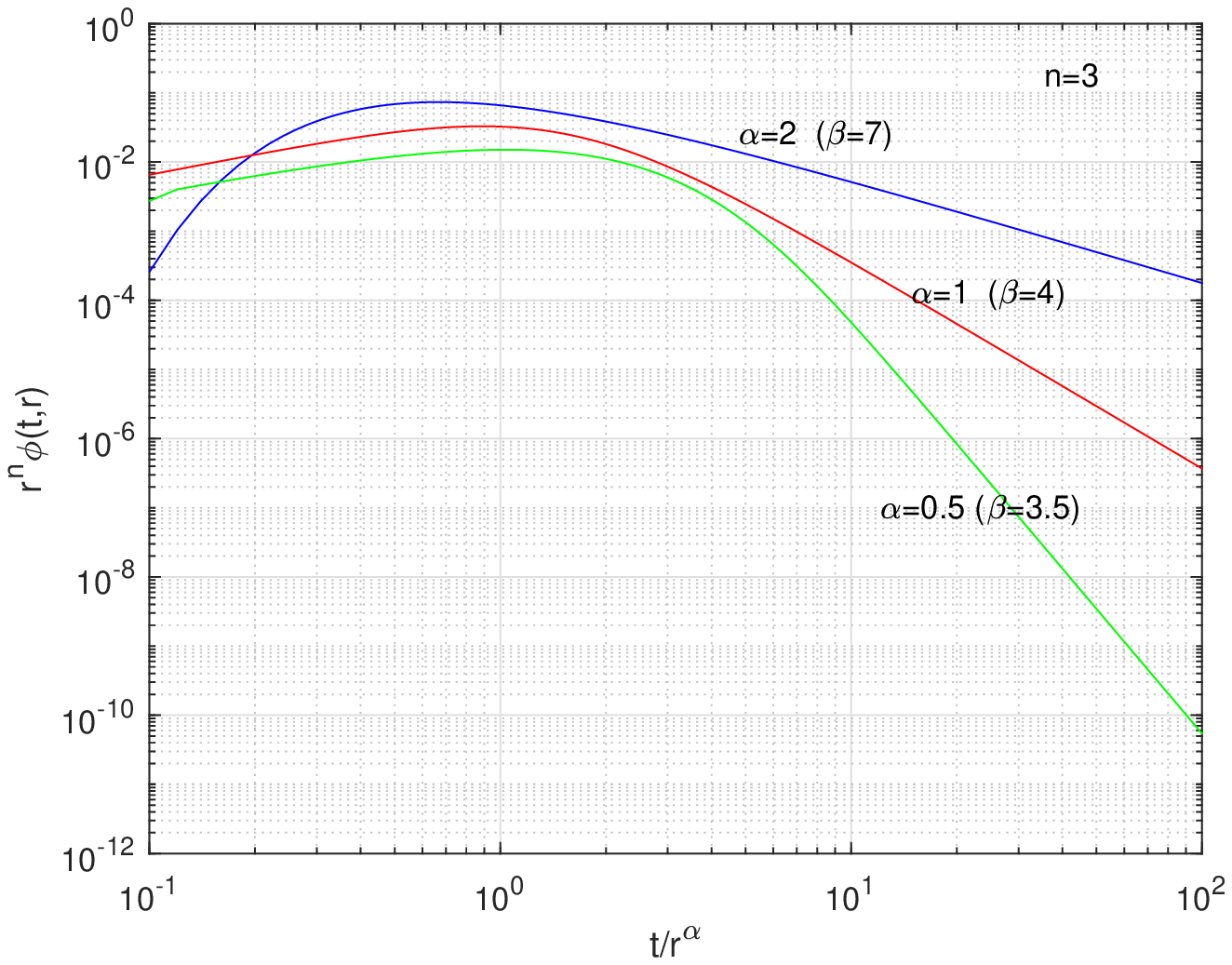}
\caption{The $r^{n}\protect\phi \left( t,r\right) $ function in dimension $3$
for the power law case}
\label{PHI_3}
\end{figure}

\begin{figure}[htb]
\centering
\includegraphics[width=0.75\textwidth]{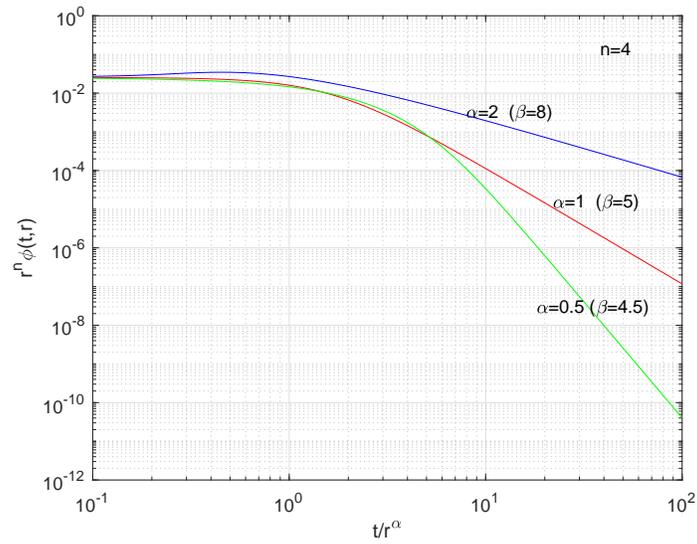}
\caption{The $r^{n}\protect%
\phi \left( t,r\right) $ function in dimension $4$ for the power law case}
\label{PHI_4}
\end{figure}

The configuration space solution $\phi _{1}\left( t,x\right) $ of (\ref{N1})
would be%
\begin{eqnarray}
\phi _{1}\left( t,r\right)  &=&\frac{1}{\left( 2\pi \right) ^{\frac{n}{2}}}%
\int_{0}^{\infty }d\left\vert k\right\vert \frac{\left\vert k\right\vert ^{%
\frac{n}{2}}}{r^{\frac{n}{2}-1}}J_{\frac{n}{2}-1}\left( \left\vert
k\right\vert r\right) e^{\left( \widetilde{\rho }\left( \left\vert
k\right\vert \right) -1\right) t}  \nonumber \\
&\simeq &\frac{1}{\left( 2\pi \right) ^{\frac{n}{2}}}\left\{
\int_{0}^{\left\vert k\right\vert _{b}}d\left\vert k\right\vert \frac{%
\left\vert k\right\vert ^{\frac{n}{2}}}{r^{\frac{n}{2}-1}}J_{\frac{n}{2}%
-1}\left( \left\vert k\right\vert r\right) e^{-A\left\vert k\right\vert
^{2}t}+\int_{\left\vert k\right\vert _{b}}^{\infty }d\left\vert k\right\vert 
\frac{\left\vert k\right\vert ^{\frac{n}{2}}}{r^{\frac{n}{2}-1}}J_{\frac{n}{2%
}-1}\left( \left\vert k\right\vert r\right) e^{-B\left\vert k\right\vert
^{\alpha }t}\right\}   \nonumber \\
&\simeq &\frac{1}{\left( 2\pi \right) ^{\frac{n}{2}}}\frac{1}{r^{n}}%
\int_{r\left\vert k\right\vert _{b}}^{\infty }dzz^{\frac{n}{2}}J_{\frac{n}{2}%
-1}\left( z\right) \exp \left( -Bz^{\alpha }\frac{t}{r^{\alpha }}\right) 
\label{N15}
\end{eqnarray}%
Therefore for times not extremely large, the breaking point $\left\vert
k\right\vert _{b}$ of the slope in $\widetilde{\rho }_{1}\left( \left\vert
k\right\vert \right) $ being very small, one concludes that $r^{n}\phi
_{1}\left( t,r\right) $ is a function of $\frac{t}{r^{\alpha }}$,%
\begin{equation}
r^{n}\phi _{1}\left( t,r\right) \simeq \frac{1}{\left( 2\pi \right) ^{\frac{n%
}{2}}}\int_{0}^{\infty }dzz^{\frac{n}{2}}J_{\frac{n}{2}-1}\left( z\right)
\exp \left( -Bz^{\alpha }\frac{t}{r^{\alpha }}\right)   \label{N16}
\end{equation}%
This function is plotted in Figs. \ref{PHI_2} \ref{PHI_3} \ref{PHI_4} for $%
n=2,3$ and $4$. Notice that the integral in (\ref{N16}) is ill defined for $%
t/r^{\alpha }=0$. It is a consequence of the initial condition of the
fundamental solution being $\phi _{1}\left( 0,x\right) =\delta ^{n}\left(
x\right) \footnote{%
In contrast to the case of the heat equation, there is in general no
regularizing effect of the singular initial condition in the nonlocal
diffusion equation, the fundamental solution being $e^{-t}\delta \left(
x\right) +\phi \left( t,x\right) $.\cite{Chasseigne}. It is the smooth $\phi
\left( t,x\right) $ part of the solution that is of interest for the study
of signal propagation in the networks.}$. Otherwise, as a function $%
t/r^{\alpha }$, it provides a complete description of the time evolution of
the fundamental solution. One notices a similar behavior for all dimensions,
whenever the same $n-\beta $ is considered.

\clearpage

\begin{figure}[htb]
\centering
\includegraphics[width=0.75\textwidth]{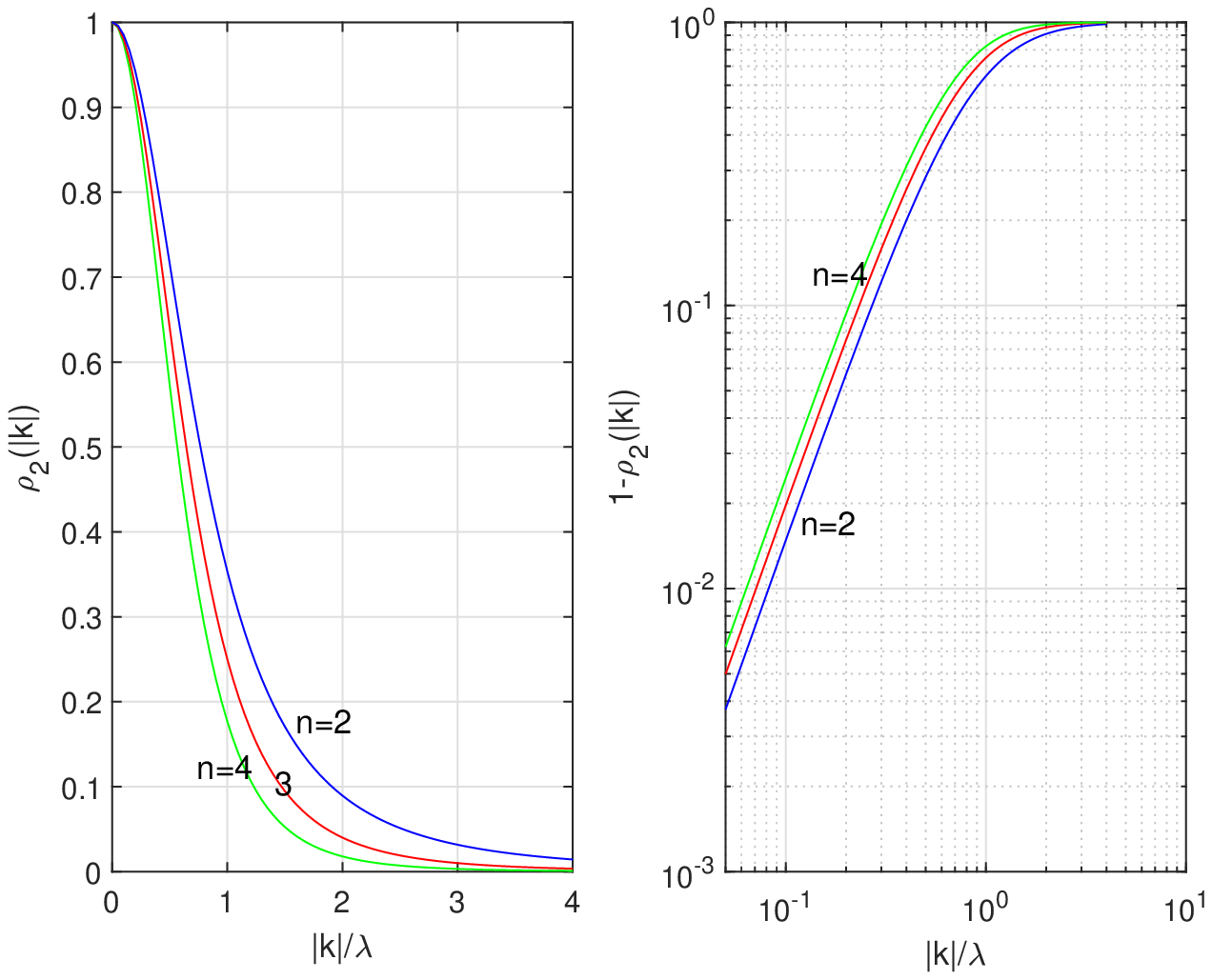}
\caption{The
Fourier kernel $\protect\rho _{2}\left( \left\vert k\right\vert \right) $
for the exponential case}
\label{EXP1}
\end{figure}

\subsection{The exponential case}

\[
\rho _{2}\left( r\right) =c_{2}e^{-\lambda r},
\]%
In this case because of the fast decay of the exponential we may, to a good
approximation, consider an infinite network. Then the normalization, $\int
d^{n}x\rho _{2}\left( r\right) =1$, implies for $n$ dimensions%
\[
c_{2}=\frac{\Gamma \left( \frac{n}{2}\right) }{\Gamma \left( n\right) }\frac{%
\lambda ^{n}}{2\pi ^{\frac{n}{2}}}
\]%
and the Fourier transform (\ref{N4}) of $\rho _{2}$ in dimension $n$ is
obtained in closed form%
\[
\widetilde{\rho }_{2}\left( \left\vert k\right\vert \right) =\frac{\lambda
^{n+1}}{\left( \lambda ^{2}+\left\vert k\right\vert ^{2}\right) ^{\frac{n+1}{%
2}}}=\frac{1}{\left( 1+\left( \frac{\left\vert k\right\vert }{\lambda }%
\right) ^{2}\right) ^{\frac{n+1}{2}}}.
\]%
and%
\begin{eqnarray*}
\phi _{2}\left( t,r\right)  &=&\frac{1}{\left( 2\pi \right) ^{\frac{n}{2}}}%
\int_{0}^{\infty }d\left\vert k\right\vert \frac{\left\vert k\right\vert ^{%
\frac{n}{2}}}{r^{\frac{n}{2}-1}}J_{\frac{n}{2}-1}\left( \left\vert
k\right\vert r\right) \exp \left( \frac{\lambda ^{n+1}}{\left( \lambda
^{2}+\left\vert k\right\vert ^{2}\right) ^{\frac{n+1}{2}}}-1\right) t \\
&=&\frac{1}{\left( 2\pi \right) ^{n}}\frac{1}{r^{n}}\int_{0}^{\infty }dzz^{%
\frac{n}{2}}J_{\frac{n}{2}-1}\left( z\right) \exp \left( \frac{1}{\left(
1+\left( z/\lambda r\right) ^{2}\right) ^{\frac{n+1}{2}}}-1\right) t
\end{eqnarray*}%
$r^{n}\phi _{2}\left( t,r\right) $ being in this case a function of $\left(
t,\lambda r\right) $. In Fig.\ref{EXP1} we have plotted the $\left\vert
k\right\vert $ dependence of $\widetilde{\rho }_{2}\left( \left\vert
k\right\vert \right) $ and $1-\widetilde{\rho }_{2}\left( \left\vert
k\right\vert \right) $.

One sees that for a large $\left\vert k\right\vert /\lambda $
region the slope of $1-\widetilde{\rho }_{2}\left( \left\vert k\right\vert
\right) $ is similar to the one of normal diffusion. Hence it is only for
very small values of $\lambda $ that one should expect anomalous diffusion
effects. This is also clear from the size of the $\left\vert k\right\vert
^{2}$ coefficient in the Taylor expansion of $\widetilde{\rho }_{2}\left(
\left\vert k\right\vert \right) $%
\[
\widetilde{\rho }_{2}\left( \left\vert k\right\vert \right) =1-\frac{n+1}{%
2\lambda ^{2}}\left\vert k\right\vert ^{2}+\cdots 
\]%
Therefore propagation of information on the network is expected to be much
less efficient than in the power law case.

\clearpage

\subsection{Comparing the propagation of information in power-law and
exponential networks}

\begin{figure}[htb]
\centering
\includegraphics[width=0.75\textwidth]{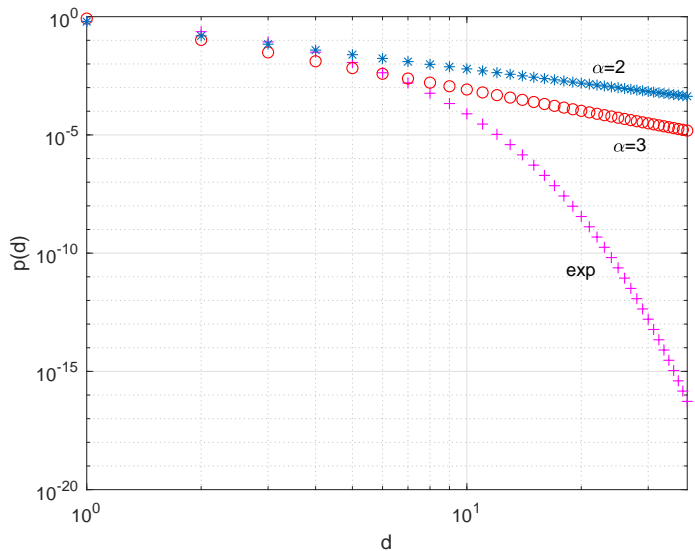}
\caption{Connection probabilities for
power law ($\protect\alpha =2$ and $3$) and exponential networks}
\label{PROPAGA_2}
\end{figure}

Here we compare by numerical simulation the propagation of information in
power law and in exponential networks. We consider a network of $40000$
agents placed on a $200\times 200$ two-dimensional lattice and establish
among them networks with unit connection strengths distributed according to
a distance-dependent law, either a power law $p\left( d\right) =\frac{%
c_{\alpha }}{d^{\alpha }}$ or $p\left( d\right) =c_{e}e^{-d}$. These
normalized probability functions are displayed in Fig.\ref{PROPAGA_2}. 

One sees that up to distance $%
\approx 10$ the connection probabilities are not very different, but for
larger distances they very much differ. The distance that is used in the
lattice is the taxi metric $d_{AB}=\left\vert x_{A}-x_{B}\right\vert
+\left\vert y_{A}-y_{B}\right\vert $.

\begin{figure}[htb]
\centering
\includegraphics[width=0.75\textwidth]{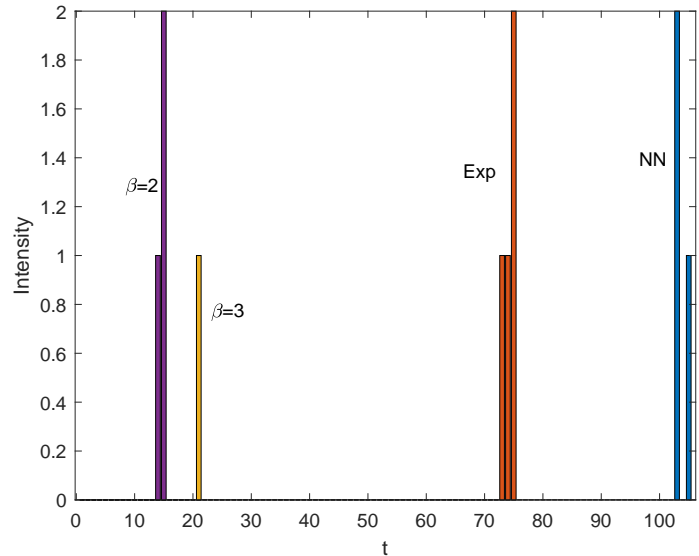}
\caption{A typical result of the
propagation of information experiment with power law ($\protect\alpha =2$
and $3$), exponential and nearest-neighbor networks}
\label{PROPAGA_1}
\end{figure}

The propagation experiments are performed in the following way: One chooses
at random two distant agents in the network, the "source" $A$ and the
"destination" $B$. At the initial time there is a unit pulse at $A$ that
this agent is going to transmit with unit intensity to all its neighbors
(meaning the agents that are connected to it, regardless of the physical
distance). At the next time step, the neighbors that received the pulse do
the same operation and so on. We use a no-cycle condition, that is, each
agent transmits the pulse to its neighbors only once, even if it receives
any other pulse later on. In this sense we are not considering exactly the
diffusion situation studied in Section 2 although the results are
qualitatively the same. The process ends at a time when the destination $B$
no longer receives any more pulses. This experiment has been repeated many
times with very similar results every time. A typical result is shown in
Fig. \ref{PROPAGA_1}. The intensities represent the number of pulses
received at each time. We have also compared the speed of propagation with a
nearest-neighbor network. All networks have the same number of connections ($%
79600$). The conclusion is that power law networks, with the same number of
connections, are extremely more efficient in the propagation of information
than the others. Although better than the nearest-neighbor one, the
exponential network fares poorly.

\textbf{Acknowledgement}: The authors acknowledge financial support from FCT
- {Funda\c{c}\~{a}o para a Ci\^{e}ncia e Tecnologia (Portugal) through
research grants UIDB/04561/2020 and UIDB/05069/2020.}


\begin{thebibliography}{99}
\bibitem{Park} H.-J. Park and K. Friston; \textit{Structural and functional
brain networks: From connections to cognition}, Science 342 (2013) 1238411.

\bibitem{Wig} G. S. Wig, B. L. Schlaggar and S. E. Petersen; \textit{%
Concepts and principles in the analysis of brain networks}, Ann. N.Y. Acad.
Sci. 1224 (2011) 126--146.

\bibitem{Knosche} T. R. Kn\"{o}sche and M. Tittgemeyer; \textit{The role of
long-range connectivity for the characterization of the
functional--anatomical organization of the cortex}, Frontiers in Systems
Neuroscience 5\ (2011) Article 58.

\bibitem{Betzel} R. F. Betzel and D. S. Bassett; \textit{Specificity and
robustness of long-distance connections in weighted, interareal connectomes}%
, PNAS 115 (2018) E4880--E4889.

\bibitem{Padula} M. C. Padula, M. Schaer, E. Scariati, A. K. Mutlu, D. Z\"{o}%
ller, M. Schneider and S. Eliez; \textit{Quantifying indices of short- and
long-range white matter connectivity at each cortical vertex}, PLOS ONE 12
(2017) 0187493.

\bibitem{Fluo} F. Drawitsch, A. Karimi, K. M. Boergens and M. Helmstaedter; 
\textit{FluoEM, virtual labeling of axons in three-dimensional electron
microscopy data for long-range connectomics}, eLife 7 (2018) e38976.

\bibitem{Barttfeld} P. Barttfeld et al.; \textit{Organization of brain
networks governed by long-range connections index autistic traits in the
general population}, Journal of Neurodevelopmental Disorders 2013, 5:16.

\bibitem{Hogan} B. Hogan; \textit{Visualizing and Interpreting Facebook
Networks} in "Analysing Social Media Networks with NodeXL", D. L. Hansen et
al. (Eds.), pp. 165-179, Elsevier 2011.

\bibitem{Romantic} C. J. Billedo, P. Kerkhof and C. Finkenauer; \textit{The
use of social networking sites for relationship maintenance in long-distance
and geographically close romantic relationships}, Cyberpsychol Behav. Soc.
Netw. 18 (2015) 152-157.

\bibitem{Carvalho} R. Carvalho and G. Iori; \textit{Socioeconomic networks
with long-range interactions}, Phys. Rev. E 78 (2008) 016110.

\bibitem{Gustafson} K. B. Gustafson, B. S. Bayati and P. A. Eckhoff; \textit{%
Fractional Diffusion Emulates a Human Mobility Network during a Simulated
Disease Outbreak}, Frontiers in Ecology and Evolution 5 (2017) Article 35.

\bibitem{Riascos1} A. P. Riascos and J. L. Mateos; \textit{Fractional
dynamics on networks: Emergence of anomalous diffusion and L\'{e}vy flights}%
, Phys. Rev. E 90 (2014) 032809.

\bibitem{Riascos2} A. P. Riascos, T.M. Michelitsch, B.A. Collet, A. F.
Nowakowski and F.C.G.A. Nicolleau; \textit{Random walks with long-range
steps generated by functions of Laplacian matrices}, J. Stat. Mech. (2018)
043404.

\bibitem{Estrada1} E. Estrada, J.-C. Delvenne, N. Hatano, J. L. Mateos, R.
Metzler, A. P. Riascos and M. T. Schaub; \textit{Random multi-hopper model:
super-fast random walks on graphs}, Journal of Complex Networks 6 (2018)
382--403.

\bibitem{Weng} T. Weng, J. Zhang, M. Khajehnejad, M. Small, R. Zheng and P.
Hui; \textit{Navigation by anomalous random walks on complex networks},
Scientific Reports 6 (2016) 37547

\bibitem{Nigris} S. de Nigris, T. Carletti and R. Lambiotte; \textit{Onset
of anomalous diffusion from local motion rules}, Phys. Rev. E 95 (2017)
022113.

\bibitem{Vilela2018} R. Vilela Mendes; \textit{Fractional networks, the new
structure}, Chaos and Complexity Letters 12 (2018) 123-128, arXiv:1804.10605.

\bibitem{Vilela2020} R. Vilela Mendes and T. Ara\'{u}jo; \textit{Long-range
connections and mixed diffusion in fractional networks}, arXiv:2002.04351.

\bibitem{LiMa} C. Li and W. Ma; \textit{Synchronizations in complex
fractional networks}, in "Handbook of Fractional Calculus with
Applications", vol. 6, I. Petr\'{a}s (Ed.) Walter de Gruyter, Berlin/Boston
2019, pp. 379-396.

\bibitem{Zuniga} C. J. Zu\~{n}iga Aguilar et al.; \textit{Fractional order
neural networks for system identification}, Chaos, Solitons and Fractals 130
(2020) 109444.

\bibitem{Rossi} F. Andreu-Vaillo, J. M. Maz\'{o}n, J. D. Rossi and J. J.
Toledo-Melero; \textit{Nonlocal diffusion problems}, Mathematical Surveys
and Monographs 165, American Math. Soc., Providence 2010.

\bibitem{Chasseigne} E. Chasseigne, M. Chaves and J. D. Rossi; \textit{%
Asymptotic behavior for nonlocal diffusion equations}, J. Math. Pures Appl.
86 (2006) 271--291.

\bibitem{Transat2020} https://www.transtats.bts.gov

\bibitem{Viana} M. P. Viana and L. F. Costa; \textit{Fast long-range
connections in transportation networks}, Phys. Lett. A 375 (2011) 1626-1629.

\bibitem{Barabasi} M. C. Gonz\'{a}lez, C. A. Hidalgo and A.-L. Barab\'{a}si; 
\textit{Understanding individual human mobility patterns}, Nature 453 (2008)
779-782.

\bibitem{Mateos} A. P. Riascos and J. L. Mateos; \textit{Networks and
long-range mobility in cities: A study of more than one billion taxi trips
in New York City}, Nature Scientific Reports 10 (2020) 4022.

\bibitem{Markov2013} N. T. Markov, M.M. Ercsey-Ravasz, C. Lamy, A.R. R.
Gomes, L. Magrou, P. Misery, P. Giroud, P. Barone, C. Dehay, Z. Toroczkai,
K. Knoblauch, D.C. Van Essen and H. Kennedy; \textit{The role of long-range
connections on the specificity of the macaque interareal cortical network},
PNAS 110 (2013) 5187-5192.

\bibitem{Markov2014} N. T. Markov et al; \textit{A Weighted and Directed
Interareal Connectivity Matrix for Macaque Cerebral Cortex}, Cerebral Cortex
24 (2014) 17--36.

\bibitem{Gamanut2017} R. Gamanut, H. Kennedy, Z. Toroczkai, D. Van Essen, K.
Knoblauch and A. Burkhalter; \textit{The Mouse Cortical Connectome
Characterized by an Ultra Dense Cortical Graph Maintains Specificity by
Distinct Connectivity Profiles}, Neuron 97 (2018) 698-715.

\bibitem{Horvat2016} S. Horvat, R. Gamanut, M. Ercsey-Ravasz, L. Magrou, B.
Gamanut, D. C. Van Essen, A. Burkhalter, K. Knoblauch, Z. Toroczkai and H.
Kennedy; \textit{Spatial Embedding and Wiring Cost Constrain the Functional
Layout of the Cortical Network of Rodents and Primates}, PLOS Biology 14
(2016) e1002512.

\bibitem{Ercsey2013} M. Ercsey-Ravasz et al; \textit{A Predictive Network
Model of Cerebral Cortical Connectivity Based on a Distance Rule}, Neuron 80
(2013) 184--197.

\bibitem{Knox} J. E. Knox et al; \textit{High-resolution data-driven model
of the mouse connectome}, Network Neuroscience 3 (2018) 217--236.

\bibitem{Ryan} R. A. Rossi and N. K. Ahmed; \textit{The Network Data
Repository with Interactive Graph Analytics and Visualization,
http://networkrepository.com}, 2015.

\bibitem{Fungi1} https://www.cs.cornell.edu/\symbol{126}%
arb/data/spatial-fungi/

\bibitem{Fungi2} S. H. Lee, M. D. Fricker and M. A. Porter; \textit{%
Mesoscale analyses of fungal networks as an approach for quantifying
phenotypic traits}, Journal of Complex Networks 5 (2017) 145--159.

\bibitem{Bucur} C. Bucur and E. Valdinoci; \textit{Nonlocal Diffusion and
Applications}, Springer, Switzerland 2016.

\bibitem{Vazquez} J. L. V\'{a}zquez; \textit{The mathematical theories of
diffusion. Nonlinear and fractional diffusion}, in Nonlocal and Nonlinear
Diffusions and Interactions: New Methods and Directions pp 205-278, Lecture
Notes in Mathematics 2186, Springer 2017.

\bibitem{Ignat-1} L. I. Ignat and J. D. Rossi; \textit{Refined asymptotic
expansions for nonlocal diffusion equations}, Journal of Evolution Equations
8 (2008) 617-629.

\bibitem{Ignat-2} L. I. Ignat and J. D. Rossi; \textit{Asymptotic behaviour
for a nonlocal diffusion equation on a lattice}, Z. Angew. Math. Phys. 59
(2008) 918-925.
\end{thebibliography}
\end{document}